\documentclass[dvips]{article}

\usepackage{icrc2011}

\newcommand{\eh}[1]{\,\mathrm{#1}}

\newcommand{\dg}{^{\circ}}

\newcommand{\degr}{$\dg$}

\newcommand{\pcnt}{$\eh{\%}$}

\newcommand{\mr}[1]{\mathrm{#1}}

\renewcommand{\epsilon}{\varepsilon}

\newcommand{\tss}[1]{\textsuperscript{#1}}

\newcommand{\tin}[1]{_{\mr{#1}}}

\title{Stereoscopic Observations of the Blazar 3C 66A with the MAGIC Telescopes}

\newcommand{\etal}{\MakeLowercase{\textit{et al. }}} 
\shorttitle{Klepser \etal Stereoscopic Observations of 3C 66A}

\authors{Stefan Klepser$^{1}$, Koji Saito$^{2}$ for
the MAGIC collaboration}
\afiliations{$^1$IFAE, Edifici Cn., Campus UAB, E-08193 Bellaterra, Spain\\
$^2$Max-Planck-Institut f\"ur Physik, D-80805 M\"unchen,
Germany}
\email{klepser@ifae.es}

\abstract{3C 66A is an intermediate-frequency peaked BL Lacertae object of
uncertain redshift. We report recent observations of the region around the blazar with the MAGIC
telescopes. The source was observed and detected in 2009 December and 2010
January, in $2.3\eh{h}$ of
good quality data. The signal could clearly be assigned to the blazar
3C 66A, statistically and systematically rejecting the nearby radio
galaxy 3C 66B as a possible origin of the gamma-ray signal by 3.6
standard deviations. The derived integral flux above $100\eh{GeV}$ is
8.3\pcnt\ 
of the Crab Nebula flux, and the energy spectrum is reproduced by a
power law of photon index $3.64\pm0.39\mathrm{(stat.)}\pm0.25\mathrm{(sys.)}$. Within the errors,
this is compatible with the spectrum derived by VERITAS in 2009. From
the spectra corrected for absorption by the extragalactic background
light, we only find small differences between the four modellings that
we applied, and constrain the redshift of the blazar to $z < 0.68$.}

\begin{document}
\maketitle

\section{Introduction}

3C~66A is a very high energy (VHE) gamma-ray blazar classified as a
BL Lac object \cite{mac87}, or an intermediate-frequency peaked BL Lac
object (IBL, \cite{per03}). It has an uncertain redshift reported to
be 0.444 \cite{mil78,lan93} or 0.321 \cite{wur96}, estimated from a single
spectral line, or a marginally resolved host galaxy respectively. Besides that, the existing upper and
lower limits to the redshift are $>0.096$~\cite{fin08}, $<0.44$~\cite{pra10}
and $<0.58$~\cite{yan10b}. 

In the VHE band, the source was first claimed by the Crimean
Astrophysical Observatory above $900\eh{GeV}$
with an integral
flux of $(3\pm1)\times 10^{-11}\eh{cm^{-2} s^{-1}}$ \cite{ste02}.
Later
observations by HEGRA and Whipple reported upper limits \cite{aha00, hor04}, while
STACEE found a hint of a signal at
a significance level of 2.2~\cite{bra05}.
More recent VERITAS observations of 3C~66A took place in 2007 and 2008, for a
total of 32.8 hours, and resulted 
in a clear detection in VHE gamma rays \cite{acc09}.
The derived energy spectrum was compatible with a power law of photon index
$\Gamma = 4.1\pm0.4_{\rm stat}\pm0.6_{\rm sys}$ and an integral flux above
$200\eh{GeV}$ of 
($1.3\pm 0.1)\times 10^{-11}\eh{cm^{-2}\,s^{-1}}$ (6\pcnt\ of the Crab Nebula flux).  

In the GeV band, the gamma-ray emission spot 3EGJ0222+4253 measured by EGRET
was associated to 3C~66A, although an influence by the 
nearby pulsar PSR J0218+4232 could not be excluded \cite{har99, kui00}.
Fermi/LAT has monitored 3C~66A since 2008 August, covering the latter
part of the VERITAS observation. According to
\cite{abd09}, which reported 
the first 5.5 months of Fermi/LAT data of 
3C~66A, it showed a significant flux variability (a factor of 5-6
between the highest and lowest fluxes). In the new multiwavelength study
published in \cite{abd11}, photon indices for the dark period
($1.9\pm0.1\tin{stat}\pm0.1\tin{sys}$) and flare
period ($1.8\pm0.1\tin{stat}\pm0.1\tin{sys}$) were estimated. 
In combination with the VERITAS spectrum, this indicates a softening of the spectrum above $100\eh{GeV}$.

From MAGIC observations of that sky region in 2007, we reported 
a significant VHE gamma-ray signal centered at
2\tss{h}23\tss{m}12\tss{s}, 43$^\circ$0$^\prime$7$''$.
This excess (named MAGIC~J0223+430) coincides within uncertainties with the position of a nearby, 
Fanaroff-Riley-I (FRI) type galaxy 3C~66B \cite[$z=0.0215$,]{stu75}. Still,
judging from the skyplot alone, and taking into account statistical and
systematic errors, the probability of the emission to originate
from 3C~66A was 14.6\pcnt.
The energy spectrum of MAGIC~J0223+430
was compatible with a power-law with an index of $\Gamma = 3.1\pm0.3$.
The integral flux above $150\eh{GeV}$ corresponded to 
($7.3\pm 1.5)\times 10^{-12}\eh{cm^{-2}\,s^{-1}}$ (2.2\pcnt\ of the Crab Nebula flux).
According to \cite{tav08}, the radio galaxy is also a plausible source of VHE 
gamma-ray radiation. Also, the recent MAGIC detection of IC
310 \cite{ale10b}, a radio galaxy at a very similar redshift ($z=0.0189$)
indicates that 3C~66B is in principle a feasible object to explain all or part of the MAGIC detection from 2007.

\section{Data set and analysis methods}

In August 2009, 3C~66A went into an optical high state which was 
reported by the Tuorla blazar monitoring
program\footnote{http://users.utu.fi/kani/1m/index.html}. This outburst
triggered new observations by
the MAGIC telescopes, located on
the Canary Island of La Palma (28.8\degr~N, 17.8\degr~W, $2220\eh{m\,a.s.l.}$). The
two $17\eh{m}$ diameter telescopes use the atmospheric Cherenkov imaging technique and
allow for gamma-ray measurements at a threshold as low as $50\eh{GeV}$ in normal trigger
mode. 

We observed 3C~66A in several time slots between 2009 September and
2010 January. However, the "starguider" CCD cameras that are used to
cross-check the telescope pointing only became fully applicable to stereo
observations in early December. To allow for a high-confidence statement on
the directional origin of the gamma rays, we only used data taken after these upgrades, which
were 5.6 hours in total. Furthermore, we had to discard data with low event
rates, affected by the exceptionally bad weather conditions in that winter.
Finally, we had 2.3 hours of good quality data left after all quality cuts.
They were taken on 6 days
between 2009 December 5 and 2010 January 18, partly under low intensity moon
light conditions.

The data were taken using the so-called "wobble" method \cite{fom94}, in which the pointing
direction alternates every 20 minutes between two positions, offset by $\pm0.4\dg$ in RA from
the source. These wobble positions were chosen to be centered on 3C~66A, but the small distance to
3C~66B (0.01\degr) allows equal judgment for both sources. The data were taken
at a zenith distance
between 13\degr\ and 35\degr. 


The analysis we present here involved only events that were triggered by both
MAGIC telescopes. The analysis was done with the MARS analysis framework
\cite{mor09}, taking advantage both of the advanced single-telescope
algorithms (e.g. \cite{ali09a}) and newly developed stereoscopic analysis
routines. A paper about these stereoscopic analysis methods of MAGIC is in
preparation.

In the skymapping procedure of MAGIC, which is crucial for the directional
statements, first an exposure model is calculated in coordinates relative to the viewing
direction. From that, a background expectation
distribution is sampled randomly to celestial coordinates. After applying
a folding with a Gaussian kernel, the background and real event distribution
are
compared using equation (17) of
\cite{lim83} as a test statistic. The resulting null-hypothesis distribution
of this test statistic is mostly resembling a gaussian distribution, but in
some cases has poissonian components as well, therefore the detection
significance is always taken from the (unsmeared, and unmodeled) $\theta^2$-distributions.
The performance of the analysis software was checked with
contemporaneous Crab Nebula data and simulations. The achieved angular
resolution\footnote{defined as
the $\sigma$ of a two-dimensional
Gaussian function}.
was 0.1\degr\ at $100\eh{GeV}$ and approaching 0.065\degr\ at
higher energies.

The systematic uncertainty on the
direction reconstruction is a product of the telescope pointing uncertainty
and possible biases that occur in the reconstruction algorithms in the
presence of e.g. star light inhomogeneities or electronic imperfections in the
hardware.
Both the total
pointing deviation and the telescope pointing precision of MAGIC were always
monitored over the years \cite{bre09, ale10}, and along with studies of
contemporary stereo data of known direction lead to an estimate of the maximal systematic
stereoscopic pointing uncertainty of 0.025\degr.

The
publicly accessible Fermi/LAT
data\footnote{http://fermi.gsfc.nasa.gov/} were analyzed using the public 
software package LAT Science Tools v9.15.2, including the Instrument Response File 
P6\_V3\_DIFFUSE, and galactic, extragalactic and instrumental background models.

\section{Results}

\begin{figure}[!t]
  \vspace{5mm}
  \centering
  \includegraphics[width=2.5in]{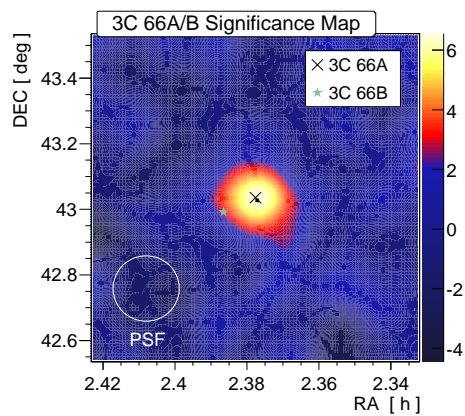}
  \caption{MAGIC significance test statistic skymap of the region around 3C~66A/B for events
with reconstructed energies above $100\eh{GeV}$.}
  \label{fig1}
 \end{figure}

\begin{figure}[!t]
  \vspace{5mm}
  \centering
  \includegraphics[width=2.5in]{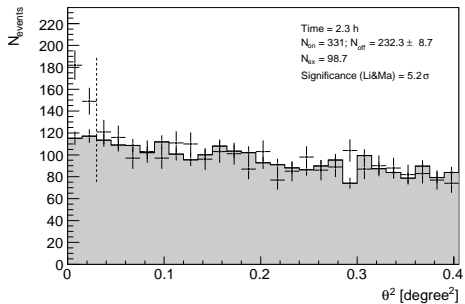}
  \caption{$\theta^2$ distribution with respect to the position of 3C~66A
($\theta^2$) for events with reconstructed energies
above $100\eh{GeV}$. The OFF data are taken from three
positions that are symmetrical with respect to the telescope pointing
directions.}
  \label{fig2}
 \end{figure}

A a skymap of the observed region above $100\eh{GeV}$ is shown in
Fig.~\ref{fig1}.
To calculate a conservative detection significance, we investigated the
distribution of squared angular distances ($\theta^2$) between photon
directions and the assumed source position.
The expected background is extracted from the opposite side of the field of
view, at the same
distance from the pointing direction. Comparing the events at the source
position with this expectation 
we find a significance of $5.2\eh{\sigma}$ (see Fig.~\ref{fig2}).

Unlike in the previous 2007 detection of MAGIC, the emission peak this time is
clearly located on top of 3C~66A. The fitted center of gravity
of the excess (small black square in Figure~\ref{fig1}) is at a distance of
$0.010\dg \pm 0.023\dg\tin{stat} \pm 0.025\dg\tin{sys}$
from 3C~66A, and $0.108\dg \pm 0.023\dg\tin{stat} \pm
0.025\dg\tin{sys}$ from 3C~66B. While being compatible with the
blazar 3C~66A,
the statistical rejection power for the emission to emerge from the radio
galaxy 3C~66B corresponds to 4.6 standard deviations. Even considering the unlikely case
of a systematic offset exactly towards the blazar, the
rejection significance of 3C~66B is $3.6\eh{\sigma}$. 

To derive an energy spectrum, we compared four different unfolding
algorithms \cite{alb07} which correct for efficiency, smearing and biasing
effects in the energy response of the detector. Among these, also the so-called \textit{forward
unfolding} was tested, which essentially is a fit with correlations defined by
the response matrix. With all unfolding methods, we found that the data are well compatible with a
power law of the form

\begin{equation}
\frac{dF}{dE} = K_{200}\left(\frac{E}{200\,\mathrm{GeV}}\right)^{-\Gamma}.
\end{equation}

\begin{figure}[!t]
  \vspace{5mm}
  \centering
  \includegraphics[width=3.in]{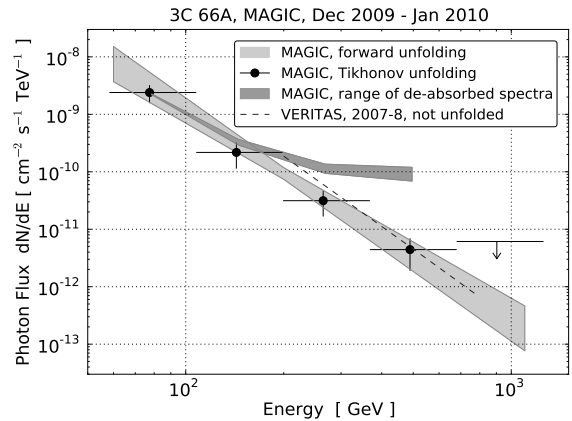}
  \caption{Differential energy spectra of 3C~66A in the period of 2009 Dec and 2010 Jan. The
light-shaded area indicates the $1\eh{\sigma}$ range of the observed unfolded power law spectrum
gained by forward unfolding, the
crosses are from the unfolding after \cite{tik79} for comparison. The
dark-shaded area is the spread of the EBL-corrected, mean flux values obtained by the
four applied EBL models, assuming the redshift of z=0.444. The VERITAS
(observed) spectrum after \cite{acc09} is shown for comparison.
}
  \label{fig3}
 \end{figure}

Here the photon index is $\Gamma = 3.64 \pm 0.39_{\rm stat} \pm 0.25_{\rm sys}$ and
the flux constant $200\eh{GeV}$ of $K_{200} = 9.6 \pm 2.5_{\rm stat} \pm
3.4_{\rm sys}\, \times 10^{-11}\eh{cm^{-2}\,s^{-1}\,TeV^{-1}}$. 
The integral flux above $100\eh{GeV}$ is equivalent to $(4.5\pm1.1)\times
10^{-11}\eh{cm^{-2}\,s^{-1}}$ 
(8.3\pcnt\ Crab Nebula flux). The
parameters and statistical errors are taken from the forward unfolding,
while the systematic errors reflect the variations among the other
unfolding algorithms, plus several additional uncertainties discussed
in \cite{alb08a}. The systematic flux uncertainties add up to about 36\pcnt\ in
total. Figure~\ref{fig3} displays the spectral points derived using the
Tikhonov unfolding method
\cite{tik79}, and the function we fitted through forward
unfolding.

We also analyzed the Fermi data from the same time period. We found the flux
variability to be insignificant in a week-to-week lightcurve.
Given the statistical uncertainties of the lightcurve, we would be
sensitive on $3\eh{\sigma}$ level to flux variations of 60\pcnt\ or greater, and
conclude the variability in the days we observed must be less than
that. The averaged flux above $200\eh{MeV}$ is comparable to the
one found in \cite{abd09}, and less than the flux level seen
in 2008 October. A
single power-law model can reproduce the source spectrum, and the photon
index is again compatible with the one found in \cite{abd11}, indicating no
significant change in the overall spectral shape.

\section{Discussion and Conclusions}

MAGIC detected a VHE gamma-ray signal from 3C~66A in the period between
December 2009 and January 2010, and during
an 
optical active state of 3C~66A~\cite{ale11}.
We rule out the emission to come
from 3C~66B at a confidence level of $3.6\eh{\sigma}$. In our previous detection
from 2007 we could not significantly reject any of the two sources, so the new
detection does not contradict the previous conclusion that the signal in 2007
might have been emerging from the radio galaxy. In particular, both objects
are not only viable VHE emitters, but both would be likely to be variable
sources.

Given the strong flux we detected in 3C~66A, we furthermore conclude that 3C~66A might have to
be in a low flux state in order not to outshine the comparably weak emission
from 3C~66B, assuming it is responsible for the flux level MAGIC detected in 2007.

The energy spectrum we unfolded is softer than in the previous MAGIC detection
($\Gamma = 3.10 \pm 0.31_{\rm stat} \pm 0.2_{\rm sys}$), and
at the same time compatible with the VERITAS spectrum of 3C~66A.
Compared to VERITAS, MAGIC has a lower threshold and the spectrum is 
extending to well below $100\eh{GeV}$. The flux level of 8.3\pcnt\ Crab Nebula
flux is similar to the one reported by VERITAS (6\pcnt), and significantly
higher than in the previous MAGIC observation (2.2\pcnt).

The VHE gamma rays produced at the source can be absorbed in the intergalactic
space by pair production with the UV to infrared photons of extragalactic background light (EBL)
\cite{ste92,hau01}. The amount of absorption depends on the energy and
redshift, and can be corrected for in the data, assuming a given modeling of the
EBL density. This kind of \textit{de-absorbed spectrum} can be regarded as the
source-intrinsic spectrum that we would measure if there were no EBL. To
derive such a de-absorbed spectrum,
we tested several state of the art EBL models, namely \cite{fra08}, the fiducial model in \cite{gil09}, \cite{kne10} and \cite{dom10}. 
The EBL corrections were applied in the unfolding procedure, because it needs the covariance matrix to correctly calculate the errors. 
The spread of the differential, de-absorbed flux spectra, obtained with the
four models and assuming the redshift of $z=0.444$, is shown as the dark
shaded area in Figure~\ref{fig3}. 
The de-absorbed photon indices for the four EBL
modelings we used are $2.57\pm0.68$ \cite{fra08}, $2.61\pm0.67$ \cite{gil09},
$2.59\pm0.68$ \cite{dom10} and $2.37\pm0.70$ \cite{kne10}. We conclude that
the differences between the de-absorbed spectra are very small, reflecting the fact that also the predicted EBL shapes and
densities are very
similar. 



Following the predicions of most VHE emission models, the de-absorbed spectrum
of a blazar is not expected to be concave, i.e. rising towards higher energies.
Two ways of testing this are to comparing the various
points of our own spectrum, or to compare the points with the Fermi photon
index. The fact that we find our spectrum neither significantly
concave, nor harder than in Fermi, suggests that the assumed redshift of
$z=0.444$ does not contradict our observations.
We investigated the plausibility of the redshift quantitatively,
assuming that the intrinsic spectrum is not expected to be exponentially
rising, and thus have a \textit{pile-up}, at highest
energies. This common method was previously used and described for example in \cite{maz07a, maz07b}.
Using the \cite{fra08} model and the likelihood ratio test between the "power
law" and "power law + pile-up" hypotheses, as described in the reference, we derive an upper limit on the redshift of 
$z<0.68$. 

Achieving these results in only $2.3\eh{h}$ of observations demonstrates the
striking
advantages of the MAGIC stereoscopic system with respect to its monoscopic era. Further MAGIC and other
gamma-ray observations of this region can provide interesting information
about the IBL type BL Lac object 3C~66A, and, during low flux periods of that,
also the FRI type galaxy 3C~66B.

\clearpage

\end{document}